\documentclass[12pt]{article}
\usepackage{amsmath}
\usepackage{graphicx}
\usepackage{authblk} % Add this package for affiliations

\usepackage[utf8]{inputenc}
\usepackage[numbers, compress]{natbib}
\usepackage{natbib}
\usepackage{slashed}
\usepackage{epsfig}
\usepackage{amsfonts}

\title{T-Duality Effects in Electrodynamics: The $(2+1)$-dimensional Case}

\author[1]{Patricio Gaete}
\author[2,3,4]{Piero Nicolini}
\affil[1]{Departamento de Física, Centro Científico-Tecnológico de Valparaíso (CCTVal), Universidad Técnica Federico Santa María, Valparaíso, Chile}
\affil[2]{Dipartimento di Fisica, Università degli Studi di Trieste, Strada Costiera, 11, 34151 Trieste TS, Italy}
\affil[3]{Istituto Nazionale di Fisica Nucleare (INFN), Sezione di Trieste, Via Alfonso Valerio, 2, 34127 Trieste TS, Italy} 
\affil[4]{Institut für Theoretische Physik, Johann Wolfgang Goethe-Universität, Max-von-Laue-Str. 1, 60438 Frankfurt am Main, Germany}

\date{\today} % Keeps the date as today

\usepackage[draft,colorinlistoftodos,shadow, textwidth=3cm, textsize=scriptsize]{todonotes}

\begin{document}

\maketitle
\begin{abstract}
We investigate the interplay between T-duality and (2+1)- dimensional electrodynamics, revealing a relationship between short and large length scales of the gauge potential. Our findings demonstrate that the electrostatic potential energy between static charges is no longer divergent at short distances in the presence of T-duality effects. It remains logarithmic at large distances, suggesting the possibility of a regulatory role for the T-duality scale \( l_0 \) in the space where the radial coordinate goes into its inverse. We also discuss the potential of T-duality to elucidate fractalization effects in physical systems, paving the way for future research on the implications for superconductors and condensed matter systems in general.\end{abstract}

\section{Introduction}

In recent years, the study of \((2+1)\)-dimensional electrodynamics has garnered increasing attention due to its unique properties and implications for theoretical physics. We briefly recall that, unlike its 3+1-dimensional counterpart, \((2+1)\)-dimensional Yang-Mills theories are super-renormalizable, with mass addition to gauge fields preserving gauge symmetry \cite{Deser:1981wh,Deser:1982vy}, while Yang-Mills-Chern-Simons models exhibit ultraviolet finiteness \cite{Barnich:1998ke}. Recently, three-dimensional models have been linked to brane dynamics \cite{Singh:2007vw} and mechanisms for supersymmetry breaking \cite{Bergman:1999na}. Furthermore, \((2+1)\)-dimensional theories explain a rich phenomenology, including high-temperature superconductivity \cite{Das:1997gg} and bosonic collective excitations in some low-dimensional condensed matter systems \cite{Khveshchenko:2001zza,Khveshchenko:2001zz}. Additionally, they alleviate the computational complexity of lattice calculations for strongly correlated systems \cite{Feuchter:2007mq}.

This paper explores the interplay between T-duality, a pivotal concept originating from string theory \cite{Sathiapalan:1986zb}, and  \((2+1)\)-dimensional electrodynamics. 
In a broad sense, T-duality can be used to relate two physical theories: one defined on a compact dimension of size $ R $ and the other on a compact dimension of size $1/R$. 
Furthermore, T-duality is intrinsically linked to specific gravity's behavior at extremely high energies.
As many authors have noted  \cite{Amati:1988tn,Maggiore:1993rv,Garay:1994en,Kempf:1994su,Adler:2001vs,Aurilia:2013psa,Dvali:2010jz,Carr:2014mya,Carr:2015nqa}, probing infinitely small length scales is impossible, regardless of the energy available. In fact, an attempt to compress matter, illuminate regions, or collide particles at length scales below the Planck length results in the formation of a black hole. A further increase in energy corresponds to an increase in the black hole's mass and size. This behavior contrasts with expectations from the Compton wavelength.
From this perspective, length scales below the Planck mass and those above it exhibit a symmetry relation reminiscent of T-duality.

The gauge field in 2+1 dimensions also exhibits a relation between short and large length scales.
In 3+1 dimensions, the potential $A_0$ diverges at short scales but remains finite at large scales. 
Conversely, in 1+1 dimensions, it diverges at large scales and is finite at short scales.
In $2+1$ dimensions, we have 
\begin{equation}
A_0 \propto \ln(\mu r),
\label{eq:amumaxwell}
\end{equation}
which diverges at both short distances ($r \ll \mu^{-1}$) and large distances ($r \gg \mu^{-1}$), where $\mu$ is a generic mass scale.
In other words, the gauge field undergoes a non-analytic ``phase transition'' in $2+1$ dimensions between lower dimensional and higher dimensional functional dependence \cite{Mureika:2012fq,Tzikas:2018wzd,Dialektopoulos:2020ekr,Gaete:2020gdt}.

This paper addresses the general question of the possible repercussions of T-duality effects on the singular behavior of the gauge field in $ 2+1 $ dimensions. Despite the challenges of observing T-duality effects and Planckian phenomena, this study may significantly impact \((2+1)\)-dimensional condensed matter systems, particularly those exhibiting a strange metal phase \cite{LaNave:2019mwv}.

The structure of the paper is outlined as follows: Section \ref{sec:t-dualED} offers a detailed background on 2+1-dimensional electrodynamics and T-duality. Section \ref{sec:potential} describes our methods for computing the electrostatic potential energy. In Section \ref{sec:concl}, we present our findings and discussions. Finally, Appendix \ref{sec:alt} presents an alternative derivation of the main result.

\section{T-Duality Effects in Electrodynamics}
\label{sec:t-dualED}

We start by recalling the standard quantization method of the electromagnetic field in terms of a generating functional%
\begin{equation}
Z\left[ J \right] = \int [\mathcal{D}A]\ e^{i S[A, J]}
\label{eq:genfunctional}
\end{equation}
where  $S[A, J]$ is the action for the gauge field $A_\mu$ coupled to a source $J$. In a $2+1$ dimensional setting
the action is formulated as
\begin{equation}
S[A, J]=\int d^3 x \left[\mathcal{L}_\mathrm{free}(F)-A_\mu J^\mu + \mathcal{L}_\mathrm{gauge}(A) \right]
\label{eq:action}
\end{equation}
where $\mathcal{L}_\mathrm{gauge}\equiv -\frac{1}{\alpha} \left(\partial_\mu A^\mu\right)^2$ is a gauge fixing term.
For the Maxwell theory, $\mathcal{L}_\mathrm{free}(F)$ is the conventional term containing the square of the field strength. 
In a non-local theory, however, the free Lagrangian is modified, namely
\begin{equation}
\mathcal{L}_\mathrm{free}(F)= - \frac{1}{4}F_{\mu \nu }\, {\cal{O}} F^{\mu \nu }
\end{equation}
where  $\cal{O}$  is a non-local operator which incorporates an infinite series of derivative terms $\Delta  \equiv {\partial _\mu }{\partial ^\mu }$, capturing essential high-energy physics effects that standard formulations may overlook.

Building on this framework, we examine the implications of T-duality for electrodynamics, noting that quantum gravitational effects can significantly alter our understanding of path integrals. Padmanabhan conjectured that each path has a weight invariant under the transformation \( l \to \frac{l_0^2}{l} \), where \( l \) is the path length and \( l_0 \) is the fundamental spacetime scale known as the ``zero-point length'' \cite{Padmanabhan:1996ap}. 
As a result,
the propagator for a massive scalar field in $D$-dimensional Euclidean space can be expressed on a discrete spacetime lattice of size \(\epsilon\) as
\begin{equation}
\mathcal{G}(\mathbf{R}, \epsilon) = \sum_{n=0}^{\infty} C(n, \mathbf{R}) \exp\left[-\mu(\epsilon)n\epsilon - \frac{\lambda(\epsilon)}{n \epsilon}\right],
\end{equation}
where \(\mathbf{R}\) represents the discrete spacetime displacement, \(C(n, \mathbf{R})\) is the number of paths connecting two spacetime points, \(\mu(\epsilon)\) is a mass parameter on the lattice, and \(\lambda(\epsilon)\) is a lattice parameter that arises from the aforementioned path duality. In the continuous limit, one obtains the momentum space propagator
\begin{equation}
G(\mathbf{p}) = \frac{l_0}{\sqrt{p^2 + m^2}} K_1\left(l_0 \sqrt{p^2 + m^2}\right),
\label{eq:paddypropagator}
\end{equation}
where \(K_1(x)\) is a modified Bessel function of the second kind. The above propagator approaches its standard expression for \(l_0 \sqrt{p^2 + m^2} \ll 1\) while being exponentially suppressed for \(l_0 \sqrt{p^2 + m^2} \gg 1\).

Later, Padmanabhan, along with Spallucci and collaborators, demonstrated that \eqref{eq:paddypropagator} coincides with the lowest-order string theory correction to field propagators \cite{Smailagic:2003hm,Fontanini:2005ik}. They derived this result by considering a \(4+1\)-dimensional closed string theory that exhibits duality relations \(R \to \frac{(R^*)^2}{R}\) and \(n \leftrightarrow w\), where \(R^* = \sqrt{\alpha'}\) denotes the self-dual radius, \(\alpha'\) is the Regge slope, \(n\) represents the Kaluza-Klein excitation, and \(w\) is the winding number of the closed string mode. The parameter identification is therefore \(l_0 = 2\pi R^* = 2\pi \sqrt{\alpha'}\).

Following this line of reasoning, the specific profile for \(\mathcal{O}\) in a massless theory consistent with T-duality corrections is given by \cite{Nicolini:2019irw,Gaete:2022une,Gaete:2022ukm}
\begin{equation}
\mathcal{O} = \left[ l_0 \sqrt{\Delta} \, K_1(l_0 \sqrt{\Delta}) \right]^{-1}.
\label{eq:nonlocaloperator}
\end{equation}
It is noteworthy that \eqref{eq:paddypropagator} and \eqref{eq:nonlocaloperator} are general results that hold for the case \(D=3\). Since we aim to calculate the interaction between static sources, the operator \(\Delta\) will act on spatial coordinates only, consistent with a Euclidean signature. Therefore, we can confidently implement \eqref{eq:nonlocaloperator} in \eqref{eq:action} and express the generating functional from \eqref{eq:genfunctional} as 
\begin{equation}
Z\left[ J \right] = \exp \left( { - \frac{i}{2}\int d^3 x  \ d^3 y \
J^\mu \left( x \right)D_{\mu \nu } \left( {x,y} \right)J^\nu  \left(
y \right)} \right), \label{NLMaxwell15}
\end{equation}
where $D_{\mu \nu } \left( {x} \right) = \int {\frac{{d^3
k}}{{\left( {2\pi } \right)^3 }}} D_{\mu \nu } \left( k \right)e^{ -
ikx}$ is the propagator in the Feynman gauge, and
\begin{equation}
{D_{\mu \nu }}\left( k \right) =  - \frac{1}{{{k^2}}}\left\{ {{{\cal O}^{ - 1}}\left( {{k^2}} \right){\eta _{\mu \nu }} + \left( 
{1 - {{\cal O}^{ - 1}}\left( {{k^2}} \right)} \right)\frac{{{k_\mu }{k_\nu }}}{{{k^2}}}} \right\}. 
\end{equation}
Using the expression, $Z = e^{iW\left[ J \right]}$, and 
Eq. (\ref {NLMaxwell15}), we obtain the following form for $W\left[ J \right]$ 
\begin{eqnarray}
W\left[ J \right] &=&  - \frac{1}{2}\int {\frac{{{d^3}k}}{{{{\left( {2\pi } \right)}^3}}}J_\mu ^ * \left( k \right)} \left[ { - \frac{1}{{{k^2}}}{{\cal O}^{ - 1}}\left( {{k^2}} \right){\eta _{\mu \nu }}} \right]{J_\nu }\left( k \right) \nonumber \\
 &-& \frac{1}{2}\int {\frac{{{d^3}k}}{{{{\left( {2\pi } \right)}^3}}}J_\mu ^ * \left( k \right)} \left[ { - \frac{1}{{{k^2}}}\left( {1 - {{\cal O}^{ - 1}}\left( {{k^2}} \right)} \right)\frac{{{k_\mu }{k_\nu }}}{{{k^2}}}} \right]{J_\nu }\left( k \right). %\nonumber\\
\label{NLMaxwell25}
\end{eqnarray}
Given that the external current $J^\mu (k)$ is conserved, the last term is vanishing
and we can write 
\begin{equation}
W\left[ J \right] = \frac{1}{2}\int {\frac{{{d^3}k}}{{{{\left( {2\pi } \right)}^3}}}} J_\mu ^ * \left( k \right)\left[ {\frac{{{l_0}\sqrt { - {k^2}} {K_1}\left( {{l_0}\sqrt { - {k^2}} } \right)}}{{{k^2}}}} \right]{J^\mu }\left( k \right). \label{NLMaxwell30}
\end{equation}
The above expression represents the energy associated to the quantum field due to the presence of virtual particles. Electrostatic potential energy can be seen as the interaction energy between two static sources. In this case the current has the form  $J_\mu  \left( x \right) \!= \!\!\left[ {Q\delta ^{\left( 2
\right)} \left( {{\bf x} - {\bf x}^{\left( 1 \right)} } \right) + Q^
\prime  \delta ^{\left( 2 \right)} \left( {{\bf x} - {\bf x}^{\left(
2 \right)} } \right)} \right] \! \delta _\mu ^0$, 
corresponding to charges $Q$ and $Q^\prime$ seated at ${\bf x}^{\left( 1 \right)}$ and ${\bf x}^{\left( 2 \right)}$ respectively. Eq. \eqref{NLMaxwell30}  can thus be expressed as:
\begin{equation}
V = QQ' \int \frac{{d^2k}}{{(2\pi)^2}} \frac{{l_0 \sqrt{{\mathbf{k}^2}}}}{{\mathbf{k}^2}} K_1(l_0 \sqrt{\mathbf{k}^2}) e^{i \mathbf{k} \cdot \mathbf{r}}, \label{NLMaxwell35}
\end{equation}
where $ {\bf r} \equiv {\bf x}^{\left( 1 \right)}  - {\bf x}^{\left(2 \right)}$.

\section{Calculation of the (2+1)-D static potential}
\label{sec:potential}

To integrate the above expression \eqref{NLMaxwell35} over ${\bf k}$, we take advantage of the following representation
\begin{equation}
V(r) = QQ' \int_0^\infty dx \, e^{-x}\ {\cal I}(x), \label{NLMaxwell130}
\end{equation}
with 
\begin{equation}
{\cal I}(x)=\int \frac{d^2k}{(2\pi)^2} \frac{e^{-\frac{l_0 {\bf k}^2}{4x}}}{{\bf k}^2} e^{i {\bf k} \cdot {\bf r}}.
\end{equation}
We introduce an infrared regulator $\mu$, so that the integral over ${\bf k}$ becomes \cite{Gaete:2012yu}
\begin{eqnarray}
{\cal I}(x) %&\equiv& \mathop{\lim}_{\varepsilon \to 0} \tilde{\cal I} \nonumber\\
&=& \mathop{\lim}_{\varepsilon \to 0} \left( \mu^2 \right)^{-\frac{\varepsilon}{2}} \int \frac{d^{2+\varepsilon} k}{(2\pi)^2} \frac{e^{-\frac{l_0^2}{4x} {\bf k}^2}}{\bf k^2} e^{i {\bf k} \cdot {\bf r}} \nonumber\\
&=& \mathop{\lim}_{\varepsilon \to 0} \left( \mu^2 \right)^{-\frac{\varepsilon}{2}} \left( \frac{r^2}{4} \right)^{-\frac{\varepsilon}{2}} \gamma \left( \frac{\varepsilon}{2}, \frac{r^2 x}{l_0^2} \right),
\label{NLMaxwell135}
\end{eqnarray}
where, $\gamma \left( \frac{\varepsilon}{2}, \frac{r^2 x}{l_0^2} \right)$ is the lower incomplete Gamma function.
We recall that for $\varepsilon \to 0$ the above functions have the following limits 
\begin{equation}
\gamma \left( \frac{\varepsilon}{2}, u \right) = \frac{2}{\varepsilon} \left[ u^{\frac{\varepsilon}{2}} e^{-u} + \gamma \left( 1 + \frac{\varepsilon}{2}, u \right) \right], \label{NLMaxwell140}
\end{equation}
with  $\gamma \left( 1 + \frac{\varepsilon}{2}, u \right) \to 1 - e^{-u}$ and  $u ^{\pm\frac{\varepsilon}{2}} \to 1 \pm \frac{\varepsilon}{2} \ln \left( u \right).$

\begin{figure}[h]
\begin{center}
\includegraphics[scale=1.0]{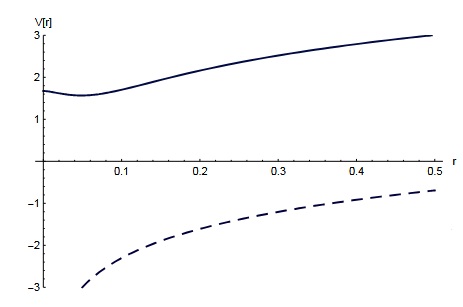}
\end{center}
\caption{\small
Shape of the modulus of the potential, $|V(r)|$ in units of $\frac{2Q^2}{\pi}$ -- see (\ref{eq:finalresult}). The dashed line represents the Coulomb potential.}
\label{fig2}
\end{figure}

Without loss of generality, we can set \( \mu \equiv \frac{1}{l_0} \) for two reasons: first, this choice does not affect the value of the field \( \mathbf{E} = -\frac{\partial V}{\partial r} \); second, there is only one physical scale in the action, namely \( l_0 \).  The above integral thus becomes
\begin{equation}
{\cal I} = -\frac{2}{\pi} \left[ \ln \left(\frac{r}{l_0}\right) - \frac{1}{2} \ln \left( \frac{r^2 x}{l_0^2} \right) e^{-\frac{r^2 x}{l_0^2}} \right], \label{NLMaxwell145}
\end{equation}
whose integral over $x$ in \eqref{NLMaxwell130} gives the final for of the potential energy, namely:
\begin{eqnarray}
{\mathrm{V}}\left({{r}}\right) &=& -\frac{2{{Q}}^{2}}{\pi}\left\{{\left({{1}
-\frac{1}{\left({{1}
+\raise0.7ex\hbox{${{{r}}^{2}}$}\!\!\left/{}\right.\!\!\lower0.7ex\hbox{${{{l}}_{0}^{2}}$}}\right)}}\right)\ln\left({\frac{{r}}{{{l}}_{0}}}\right)
+\frac{1}{\left({{1} 
+\raise0.7ex\hbox{${{{r}}^{2}}$}\!\!\left/{}\right.\,\lower0.7ex\hbox{${{{l}}_{0}^{2}}$}}\right)}\frac{C}{2}}\right\}
\nonumber\\
&+&\frac{2{{Q}}^{2}}{\pi}\frac{1}{\left({{1} +
\raise0.7ex\hbox{${{{r}}^{2}}$}\!\!\left/{}\right.\!\!\lower0.7ex\hbox{${{{l}}_{0}^{2}}$}}\right)}\ln\left({\sqrt{{1}
+
\raise0.7ex\hbox{${{{r}}^{2}}$}\!\!\left/{}\right.\!\!\lower0.7ex\hbox{${{{l}}_{0}^{2}}$}}}\right)
\label{eq:finalresult}
\end{eqnarray}
In the above result, we set $Q = -Q'$, while $C$ denotes the Euler's constant ($0.577$). In 2+1 dimensions charges have dimension $[Q]=\mathrm{mass}^{1/2}$  and $V(r)$ has the dimension of a mass as expected.  The potential energy is finite at the origin (see Fig. \ref{fig2}). Indeed for $r\to 0$ one obtains  
 \begin{equation}
    {{V}}{(}{0}{)} = -\frac{{{Q}}^{2}{C}}{\pi}
    \end{equation}
We further observe that at large distances, the potential energy diverges, as indicated by \eqref{eq:finalresult}. This divergence appears to be inescapable unless one assumes the system is confined within a finite volume. However, it is instructive to express the potential energy in terms of a new radial coordinate defined as \(\rho \equiv \frac{l_0^2}{r}\). Under this transformation, the expression from  \eqref{eq:finalresult} reveals that the potential energy becomes finite as \(\rho\) increases, but remains singular as \(\rho\) approaches zero, namely having the same form
\begin{equation}
V(\rho) \approx \frac{2 Q^2 }{\pi} \ln\left(\frac{\rho}{l_0}\right)
\end{equation}
as in the standard Maxwell case \eqref{eq:amumaxwell}. In principle, one could apply T-duality corrections in the \(\rho\)-space and obtain a potential energy that is finite at both ends. 
This observation highlights the regulating power of the parameter \(l_0\) across both regimes, depending on the definition of the spatial coordinate.
An alternative derivation of the result above is provided in Appendix A, with particular emphasis on gauge invariance, consistent with our previous work \cite{Gaete:2022une}.

\section{Conclusions}
\label{sec:concl}

The first result of this paper is the application of the T-duality argument to obtain a finite potential energy (see \eqref{eq:finalresult}). We note that, in contrast to the 3+1-dimensional case, the scale \( l_0 \) at which T-duality effects set in is arbitrary. In 2+1 dimensions, the gravitational constant is actually \( \sim l_0 \), but there is no actual experiment to constrain it. Therefore, we can set \( l_0 \) anywhere; for instance, in the nanometer to micron range. This opens up a direct connection with the phenomenology of condensed matter physics. A second result of this paper is the fractalization of space, which emerges due to a local loss of resolution, either due to the intrinsic uncertainty of quantum spacetime \cite{Ansoldi:1995by, Ansoldi:1998ys, Ambjorn:2005db, Modesto:2009qc, Nicolini:2010bj, Frassino:2013lya} or due to bulk effects in condensed matter systems \cite{Dave:2012un, Karch:2015zqd}.

We recall the definition of spacetime fractal dimension as a scaling parameter of the potential \cite{Gaete:2022une}:
\begin{equation}
\mathbb{D}\equiv 3-\frac{\partial \ln V(r)}{\partial \ln r}.
\end{equation}
In the case that \(\mathbb{D} = D\), where $D$ is the topological (discrete) spacetime dimension, the potential behaves classically. When \(\mathbb{D}\) departs from \(D\), assuming non-integer values and becoming a function of \(r\), it signals the emergence of fractal effects. To calculate \(\mathbb{D}\)—including the case \(V\sim \ln(\mu r)\)—it is convenient to represent the potential as:
\begin{equation}
V(r) \sim \int \frac{dr}{r^{\mathbb{D}-2+\varepsilon}}
\end{equation}
with \(\varepsilon > 0\) as an infinitesimal parameter. The expansion at short length scales of \eqref{eq:finalresult} gives:
\begin{equation}
V(r) \approx \frac{Q^2 C}{\pi} + \frac{V''(0)}{2} r^2 + O(r^3)
\end{equation}
which leads to:
\begin{equation}
\mathbb{D} = 3 - \frac{2\frac{V''(0)}{2} r^2}{\frac{Q^2 C}{\pi} + \frac{V''(0)}{2} r^2}.
\end{equation}
It is clear that \(\mathbb{D}\) becomes continuous, signaling fractalization. In addition, since \(V''(0) > 0\) (see Fig. \ref{fig2}), \(\mathbb{D}\) is actually smaller than the topological dimension \(D\), implying the presence of a local loss or resolution.

For the future it would be interesting to explore the repercussion of this results in the physics of superconductors  \cite{Nielsen:1973cs} or in all situation where fractal effects play a dominant role \cite{Moore91,Wen91}.

\appendix

\section{Gauge-Invariant Computation of Interaction Energy}
\label{sec:alt}

Our aim in this appendix is to recover the interaction energy as given by (\ref{eq:finalresult}), which was computed between external probe sources. However, to put our discussion into context, we begin by briefly summarizing the relevant aspects of the earlier analysis described in \cite{Gaete:2022une}. Our first undertaking is to calculate the expectation value of the Hamiltonian in the physical state $|\Phi\rangle$, denoted by $\left\langle H \right\rangle _\Phi$.

We now proceed to derive the Hamiltonian. To do so, we focus on the
Hamiltonian framework of this theory. The canonical momenta are given by ${\Pi ^\mu } = - {\cal O}{F^{0\mu }}$ and we immediately identify the sole primary constraint as $\Pi ^0=0$, then gives the canonical Hamiltonian \begin{equation}
H_\mathrm{C} = \int {{d^2}x} \left\{ { - {A_0}{\partial _i}{\Pi ^i} - \frac{1}{2}{\Pi _i}{{\cal O}^{ - 1}}{\Pi ^i} + \frac{1}{4}
{F_{ij}}{\cal O}{F^{ij}}} \right\}. \label{App-05}
\end{equation}
Time conservation of the primary constraint  $\Pi ^0$ leads to the secondary Gauss-law constraint $\Gamma_1 \left( x \right) \equiv \partial _i \Pi ^i=0$. 
The preservation of $\Gamma_1 \left( x \right) \equiv \partial _i \Pi ^i$ for all times does not lead to any more constraints. The theory is thus seen to possess only two first-class constraints. The extended Hamiltonian that generates time translations reads $H = {H_\mathrm{C}} + \int {{d^2}x\left( {{c_0}\left( x \right){\Pi _0}\left( x \right) + {c_1}\left( x \right)\Gamma_{1} \left( x \right)} \right)}$, where ${{c_0}\left( x \right)}$ and ${{c_1}\left( x \right)}$ are arbitrary Lagrange multipliers, where ${{c_0}\left( x \right)}$ and ${{c_1}\left( x \right)}$ are arbitrary Lagrange multipliers. Furthermore, it is clear that ${\dot A_0}\left( x \right) = \left[ {{A_0}\left( x \right),H} \right] = {c_0}\left( x \right)$, which is an arbitrary function. Since ${\Pi ^0} = 0$ always, neither $A^{0}$ nor ${\Pi ^0}$ is relevant for describing this system and can therefore be omitted from the theory. 

Then, the Hamiltonian takes the form
\begin{equation}
H = \int {{d^2}x} \left\{ {c\left( x \right){\partial _i}{\Pi ^i} - \frac{1}{2}{\Pi _i}{{\cal O}^{ - 1}}{\Pi ^i} + \frac{1}{4}
{F_{ij}}{\cal O}{F^{ij}}} \right\}, 
 \label{App-10}
\end{equation}
where $c(x) = c_1 (x) - A_0 (x)$.

Next, the presence of the arbitrary quantity $c(x)$ is undesirable as we have no way to assign it meaning in quantum theory. To resolve this, we introduce a supplementary condition on the vector potential, thereby rendering the complete set of constraints second class. As explained in \cite{Gaete:2022une}, a particularly convenient choice is 
\begin{equation}
\Gamma _2 \left( x \right) \equiv \int\limits_{C_{\xi x} } {dz^\nu }
A_\nu \left( z \right) \equiv \int\limits_0^1 {d\lambda x^i } A_i
\left( {\lambda x} \right) = 0,     \label{App-15}
\end{equation}
where  $\lambda$ $(0\leq \lambda\leq1)$ is the parameter describing the space-like straight path $ {z^i} = \xi ^i  + \lambda 
\left( {x -\xi } \right)^i $, and $ \xi $ is a fixed point (reference point), on a fixed time slice. There is no essential loss of generality if we 
restrict our considerations to $ \xi ^i=0 $. As a consequence, the only nontrivial Dirac bracket is given by
\begin{equation}
\left\{ {A_i \left( x \right),\Pi ^j \left( y \right)} \right\}^ *
=\delta{ _i^j} \delta ^{\left( 2 \right)} \left( {x - y} \right) 
-\partial _i^x \int\limits_0^1 {d\lambda x^j } \delta ^{\left( 2
\right)} \left( {\lambda x - y} \right). \label{App-20}
\end{equation}

We now address the problem of determining the interaction energy between point-like sources in our model. The state $|\Phi\rangle$ that represents these sources is constructed by operating creation/annihilation operators on the vacuum. We emphasize that such states are manifestly gauge-invariant by design. For our specific case, we consider the gauge-invariant string-like state $\left| \Phi  \right\rangle  \equiv \left| {\overline \Psi  \left(\bf y \right)\Psi \left( {\bf y}\prime \right)} \right\rangle$,
where a fermion is localized at ${\bf y}\prime$ an an antifermion at ${\bf y}$
as follows
\begin{equation}
\left| \Phi  \right\rangle  \equiv \left| {\overline \Psi  \left(
\bf y \right)\Psi \left( {\bf y}\prime \right)} \right\rangle 
= \overline \psi \left( \bf y \right)\exp \left(
{iQ\int\limits_{{\bf y}\prime}^{\bf y} {dz^i } A_i \left( z \right)}
\right)\psi \left({\bf y}\prime \right)\left| 0 \right\rangle,
\label{App-25}
\end{equation}
here $\left| 0 \right\rangle$ denotes the physical vacuum state. The line integral in the above expression is taken along a spacelike path starting at ${{\bf y}^ {\prime} }$ and ending at ${{\bf y}}$, on a fixed time slice, where $Q$ represents the external charge. 

Taking into account the above Hamiltonian structure, we find that $\left\langle H \right\rangle _\Phi$ is given by
\begin{equation}
\left\langle {{H_\Phi }} \right\rangle  = \left\langle \Phi  \right|\int {{d^2}x}  - \frac{1}{2}{\Pi _i}{{\cal O}^{ - 1}}{\Pi ^i}\left| \Phi  \right\rangle.  \label{App-30}
\end{equation}
By using equation (\ref{App-20}), we can rewrite the expectation value as follows
\begin{equation}
\left\langle H \right\rangle _\Phi   = \left\langle H \right\rangle
_0 + \left\langle H \right\rangle _\Phi ^{\left( 1 \right)},
\label{App-35}
\end{equation}    
where $\left\langle H \right\rangle _0  = \left\langle 0
\right|H\left| 0 \right\rangle$. Whereas the term $\left\langle H \right\rangle
_\Phi ^{\left( 1 \right)}$ is given by
\begin{eqnarray}
\left\langle H \right\rangle _\Phi ^{\left( 1 \right)} \!\! &=&\!\!
- \frac{{Q^2 }}{2}
\int {d^2 x} \int_{\bf y}^{{\bf y}^ \prime} {dz_i^ \prime}
\delta ^{\left( 2 \right)}
\left( {{\bf x} - {\bf z}^ \prime} \right) \nonumber\\
\!\!&\times&\!\!\!\left[{- {l_0}\sqrt { - \nabla _x^2} \,{K_1}\left( {{l_0}\sqrt { - \nabla _x^2} } \right)} \right]\int_{\bf y}
^{{\bf y}^ \prime}{dz_i^ \prime} 
\delta ^{\left( 2 \right)}\!\left( {{\bf x} - {\bf z}} \right),
\label{App-40}
\end{eqnarray} 
where, in this static case, $\Delta= - \nabla^2$.
The previous expression can also be written as
\begin{equation}
\left\langle H \right\rangle _\Phi ^{\left( 1 \right)} =  - \frac{{{Q^2}}}{2}\int_{\bf y}^{{{\bf y}^ \prime }} {d{z^{ \prime 
i}}} \int_{\bf y}^{{{\bf y}^ \prime }} {d{z^i}\,\nabla _z^2 \tilde G} \left( {{\bf z},{{\bf z}^ \prime }} \right),
\label{App-45}
\end{equation}
where $\tilde G$ is the Green function. % given by equation  (\ref{NLMaxwell45a}).
From the previous expression, we readily obtain the interaction energy given by expression (\ref{eq:finalresult}).

\section*{Acknowledgments}
One of us (P. G.) would like to thank the Abdus Salam ICTP for the hospitality and support. Support from INFN-Trieste is also gratefully acknowledged. Additionally, P. G. acknowledges the financial support provided by ANID PIA/APOYO AFB230003. 
P.N. wishes to thank the GNFM (Italian National Group for Mathematical Physics) and the ``Iniziativa
Specifica FLAG'' of the INFN (Italian National Institute for Nuclear Physics).


\begin{thebibliography}{10}
\def\enquote#1{``#1''}

\bibitem{Deser:1981wh}
S.~Deser, R.~Jackiw and S.~Templeton, 
%\enquote{{Topologically Massive Gauge   Theories},} 
  Annals Phys. {\bf 140}, 372 (1982), [Erratum: Annals Phys. 185,
  406 (1988)].

\bibitem{Deser:1982vy}
S.~Deser, R.~Jackiw and S.~Templeton, 
%\enquote{{Three-Dimensional Massive Gauge   Theories},} 
  Phys. Rev. Lett. {\bf 48}, 975 (1982).

\bibitem{Barnich:1998ke}
G.~Barnich,
% \enquote{{A General nonrenormalization theorem in the extended
%  antifield formalism},} 
JHEP {\bf 12}, 003 (1998).

\bibitem{Singh:2007vw}
H.~Singh, %\enquote{{3-branes on Eguchi-Hanson 6D instantons},} 
Gen. Rel. Grav.
  {\bf 39}, 839 (2007).

\bibitem{Bergman:1999na}
O.~Bergman, A.~Hanany, A.~Karch and B.~Kol, %\enquote{{Branes and supersymmetry   breaking in three-dimensional gauge theories},}
   JHEP {\bf 10}, 036 (1999).

\bibitem{Das:1997gg}
A.~K. Das, 
{\em {Finite Temperature Field Theory}\/}, World Scientific, New
  York (1997).

\bibitem{Khveshchenko:2001zza}
D.~V. Khveshchenko, %\enquote{{Magnetic field-induced insulating behavior in   highly oriented pyrolitic graphite},} 
  Phys. Rev. Lett. {\bf 87}, 206401
  (2001).

\bibitem{Khveshchenko:2001zz}
D.~V. Khveshchenko, %\enquote{{Ghost Excitonic Insulator Transition in Layered   Graphite},} 
  Phys. Rev. Lett. {\bf 87}, 246802 (2001).

\bibitem{Feuchter:2007mq}
C.~Feuchter and H.~Reinhardt, %\enquote{{The Yang-Mills vacuum in Coulomb gauge   in D=2+1 dimensions},} 
  Phys. Rev. D {\bf 77}, 085023 (2008).

\bibitem{Sathiapalan:1986zb}
B.~Sathiapalan, %\enquote{{Duality in Statistical Mechanics and String Theory},}
  Phys. Rev. Lett. {\bf 58}, 1597 (1987).

\bibitem{Amati:1988tn}
D.~Amati, M.~Ciafaloni and G.~Veneziano, %\enquote{{Can Space-Time Be Probed   Below the String Size?}} 
  Phys. Lett. B {\bf 216}, 41 (1989).

\bibitem{Maggiore:1993rv}
M.~Maggiore, %\enquote{{A Generalized uncertainty principle in quantum   gravity},} 
Phys. Lett. B {\bf 304}, 65 (1993).

\bibitem{Garay:1994en}
L.~J. Garay, %\enquote{{Quantum gravity and minimum length},} 
Int. J. Mod. Phys.
  A {\bf 10}, 145 (1995).

\bibitem{Kempf:1994su}
A.~Kempf, G.~Mangano and R.~B. Mann, %\enquote{{Hilbert space representation of
%  the minimal length uncertainty relation},} 
  Phys. Rev. D {\bf 52}, 1108
  (1995).

\bibitem{Adler:2001vs}
R.~J. Adler, P.~Chen and D.~I. Santiago, %\enquote{{The Generalized uncertainty
 % principle and black hole remnants},} 
  Gen. Rel. Grav. {\bf 33}, 2101 (2001).

\bibitem{Aurilia:2013psa}
A.~Aurilia and E.~Spallucci, \enquote{{Planck's uncertainty principle and the
  saturation of Lorentz boosts by Planckian black holes},}   (2013).

\bibitem{Dvali:2010jz}
G.~Dvali, G.~F. Giudice, C.~Gomez and A.~Kehagias, %\enquote{{UV-Completion by
  %Classicalization},} 
  JHEP {\bf 08}, 108 (2011).

\bibitem{Carr:2014mya}
B.~J. Carr, % \enquote{{The Black Hole Uncertainty Principle Correspondence},}
  Springer Proc. Phys. {\bf 170}, 159 (2016).

\bibitem{Carr:2015nqa}
B.~J. Carr, J.~Mureika and P.~Nicolini, %\enquote{{Sub-Planckian black holes and
  %the Generalized Uncertainty Principle},} 
  JHEP {\bf 07}, 052 (2015).

\bibitem{Mureika:2012fq}
J.~Mureika and P.~Nicolini, %\enquote{{Self-completeness and spontaneous
  %dimensional reduction},} 
  Eur. Phys. J. Plus {\bf 128}, 78 (2013).

\bibitem{Tzikas:2018wzd}
A.~G. Tzikas, P.~Nicolini, J.~Mureika and B.~Carr, %\enquote{{Primordial black
%  holes in a dimensionally reduced universe},} 
JCAP {\bf 12}, 033 (2018).

\bibitem{Dialektopoulos:2020ekr}
K.~F. Dialektopoulos, P.~Nicolini and A.~G. Tzikas, %\enquote{{Primordial black
 % holes in a dimensionally oxidizing Universe},} 
 JCAP {\bf 10}, 008 (2020).

\bibitem{Gaete:2020gdt}
P.~Gaete, P.~Nicolini and E.~Spallucci, %\enquote{{Regularization ambiguity and
 % van der Waals black hole in 2 + 1 dimensions},} 
  Eur. Phys. J. C {\bf 81}, 6,
  526 (2021).

\bibitem{LaNave:2019mwv}
G.~La~Nave, K.~Limtragool and P.~W. Phillips, %\enquote{{Fractional
 % Electromagnetism in Quantum Matter and High-Energy Physics},} 
 Rev. Mod. Phys.
  {\bf 91}, 2, 021003 (2019).

\bibitem{Padmanabhan:1996ap}
T.~Padmanabhan, %\enquote{{Duality and zero point length of space-time},} 
Phys.
  Rev. Lett. {\bf 78}, 1854 (1997).

\bibitem{Smailagic:2003hm}
A.~Smailagic, E.~Spallucci and T.~Padmanabhan, \enquote{{String theory T
  duality and the zero point length of space-time},}   (2003).

\bibitem{Fontanini:2005ik}
M.~Fontanini, E.~Spallucci and T.~Padmanabhan, %\enquote{{Zero-point length from
 % string fluctuations},} 
  Phys. Lett. B {\bf 633}, 627 (2006).

\bibitem{Nicolini:2019irw}
P.~Nicolini, E.~Spallucci and M.~F. Wondrak, %\enquote{{Quantum Corrected Black
 % Holes from String T-Duality},} 
 Phys. Lett. B {\bf 797}, 134888 (2019).

\bibitem{Gaete:2022une}
P.~Gaete and P.~Nicolini, %\enquote{{Finite electrodynamics from T-duality},}
  Phys. Lett. B {\bf 829}, 137100 (2022).

\bibitem{Gaete:2022ukm}
P.~Gaete, K.~Jusufi and P.~Nicolini, %\enquote{{Charged black holes from
%  T-duality},} 
Phys. Lett. B {\bf 835}, 137546 (2022).

\bibitem{Gaete:2012yu}
P.~Gaete, J.~Helayel-Neto and E.~Spallucci, %\enquote{{Aspects of finite
  %electrodynamics in D=3 dimensions},}
   J. Phys. A {\bf 45}, 215401 (2012).

\bibitem{Ansoldi:1995by}
S.~Ansoldi, A.~Aurilia and E.~Spallucci, %\enquote{{Membrane vacuum as a type II
 % superconductor},}
 Int. J. Mod. Phys. B {\bf 10}, 1695 (1996).

\bibitem{Ansoldi:1998ys}
S.~Ansoldi, A.~Aurilia and E.~Spallucci, %\enquote{{Loop quantum mechanics and
 % the fractal structure of quantum space-time},}
   Chaos Solitons Fractals {\bf
  10}, 197 (1999).

\bibitem{Ambjorn:2005db}
J.~Ambjorn, J.~Jurkiewicz and R.~Loll, %\enquote{{Spectral dimension of the
  %universe},} 
  Phys. Rev. Lett. {\bf 95}, 171301 (2005).

\bibitem{Modesto:2009qc}
L.~Modesto and P.~Nicolini, %\enquote{{Spectral dimension of a quantum
  %universe},} 
  Phys. Rev. D {\bf 81}, 104040 (2010).

\bibitem{Nicolini:2010bj}
P.~Nicolini and E.~Spallucci, %\enquote{{Un-spectral dimension and quantum
 % spacetime phases},} 
 Phys. Lett. B {\bf 695}, 290 (2011).

\bibitem{Frassino:2013lya}
A.~M. Frassino, P.~Nicolini and O.~Panella, %\enquote{{Unparticle Casimir
 % effect},} 
  Phys. Lett. B {\bf 772}, 675 (2017).

\bibitem{Dave:2012un}
K.~B. Dave, P.~W. Phillips and C.~L. Kane, %\enquote{{Absence of
 % Luttinger{\textquoteright}s theorem due to zeros in the single-particle green
 % function},}
   Phys. Rev. Lett. {\bf 110}, 9, 090403 (2013).

\bibitem{Karch:2015zqd}
A.~Karch, K.~Limtragool and P.~W. Phillips, %\enquote{{Unparticles and Anomalous
%  Dimensions in the Cuprates},} 
  JHEP {\bf 03}, 175 (2016).

\bibitem{Nielsen:1973cs}
H.~B. Nielsen and P.~Olesen, %\enquote{{Vortex Line Models for Dual Strings},}
  Nucl. Phys. B {\bf 61}, 45 (1973).

\bibitem{Moore91}
G.~{Moore} and N.~{Read}, %\enquote{{Nonabelions in the fractional quantum hall
 % effect},}
   Nuclear Physics B {\bf 360}, 2, 362 (1991).

\bibitem{Wen91}
X.~G. {Wen}, %\enquote{{Non-Abelian statistics in the fractional quantum Hall
 % states},} 
 Phys. Rev. Lett. {\bf 66}, 6, 802 (1991).

\end{thebibliography}
\end{document}